\documentclass[a4paper]{jpconf} % PAPIERFORMAT, STANDARDSCHRIFTGRÖSSE, DOKUMENTENART
\usepackage{eurosym} % ZUR VERWENDUNG DES EURO- SYMBOLS
\usepackage{amsmath} % FÜR MATHEMATISCHE SYMBOLIK
\usepackage{amssymb} % FÜR MATHEMATISCHE SYMBOLIK
\usepackage{dsfont} % FÜR MATHEMATISCHE SYMBOLIK
\usepackage{enumerate} % FÜR AUFZÄHLUNGEN

\usepackage{graphicx} % ZUM GRAPHIKEN EINBINDEN

\usepackage[mathscr]{eucal} % FÜR MATHEMATISCHE SYMBOLIK
\usepackage{longtable} % ZUM ERSTELLEN VON SEHR LANGEN TABELLEN
\usepackage{multirow}

\newcommand{\dd}{\mathrm{d}}
\newcommand{\Dp}{\partial}
\newcommand{\un}{\infty}

\newcommand{\Ra}{\Rightarrow}

\newcommand{\li}{\left}
\newcommand{\ri}{\right}

\newcommand{\cen}[1]{\begin{center} #1 \end{center}}

\begin{document}
\title{Holographic vector mesons in a dilaton background}

\author{R. Z\"ollner, B. K\"ampfer}
\address{Helmholtz-Zentrum Dresden-Rossendorf, PF 510119, D-01314 Dresden, Germany \\ and \\ Institut f\"ur Theoretische Physik, TU Dresden, D-01062 Dresden, Germany}

\begin{abstract}
Within a holographic framework, we consider vector mesons riding on a gravity-dilaton background. The latter one is determined directly from a Schr\"odinger equivalent potential which delivers a proper $\rho$ meson Regge trajectory. The mapping on the dilaton potential yields a thermodynamic phase structure with a first-order transition.
\end{abstract}

\section{Introduction}
Vector mesons couple directly to dileptons which are interesting as penetrating probes in relativistic heavy-ion collisions, in particular in a range of beam energies, where transiently large net baryon densities are achieved \cite{tetyana}. Basically, in-medium vector meson properties are accessible via spectral functions in a heat bath from QCD. But the evaluation of QCD as theory of strong interaction in the non-pertubative regime faces some problems, e.g.~the sign problem at non-zero chemical potential $\mu$ or real-time phenomena hampered by the Euclidean action formulation. One alternative is to employ suitable models which catch some QCD properties. Among such models is the holographic approach which aims at mapping the four-dimensional QCD living in a Minkowski space to a five-dimensional gravitational theory living in an asymptotic anti-de Sitter (AdS) spacetime. Here, we employ the AdS/QCD correspondence \cite{Erdmenger, Kir1, Kir2}, conjectured to emerge from its root - the AdS/CFT correspondence -, and explore a set-up, where the Regge trajectory of vector mesons serves as input (section 2). We translate that input into a dilaton potential by exploiting Einstein equations (section 3) to fix the gravity-dilaton background, where the vector meson modes are considered in the probe limit. The gravity-dilaton dynamics adjusted in such a way needs modifications to accommodate the QCD thermodynamics on the same footing. This finding seems to require either the treatment of hadrons beyond the probe limit or/and the inclusion of further fields beyond the one-component dilaton (section 4).

\section{Holographic vector mesons}
  \begin{figure}
   \cen{\parbox{9.cm}{\includegraphics[scale=.45]{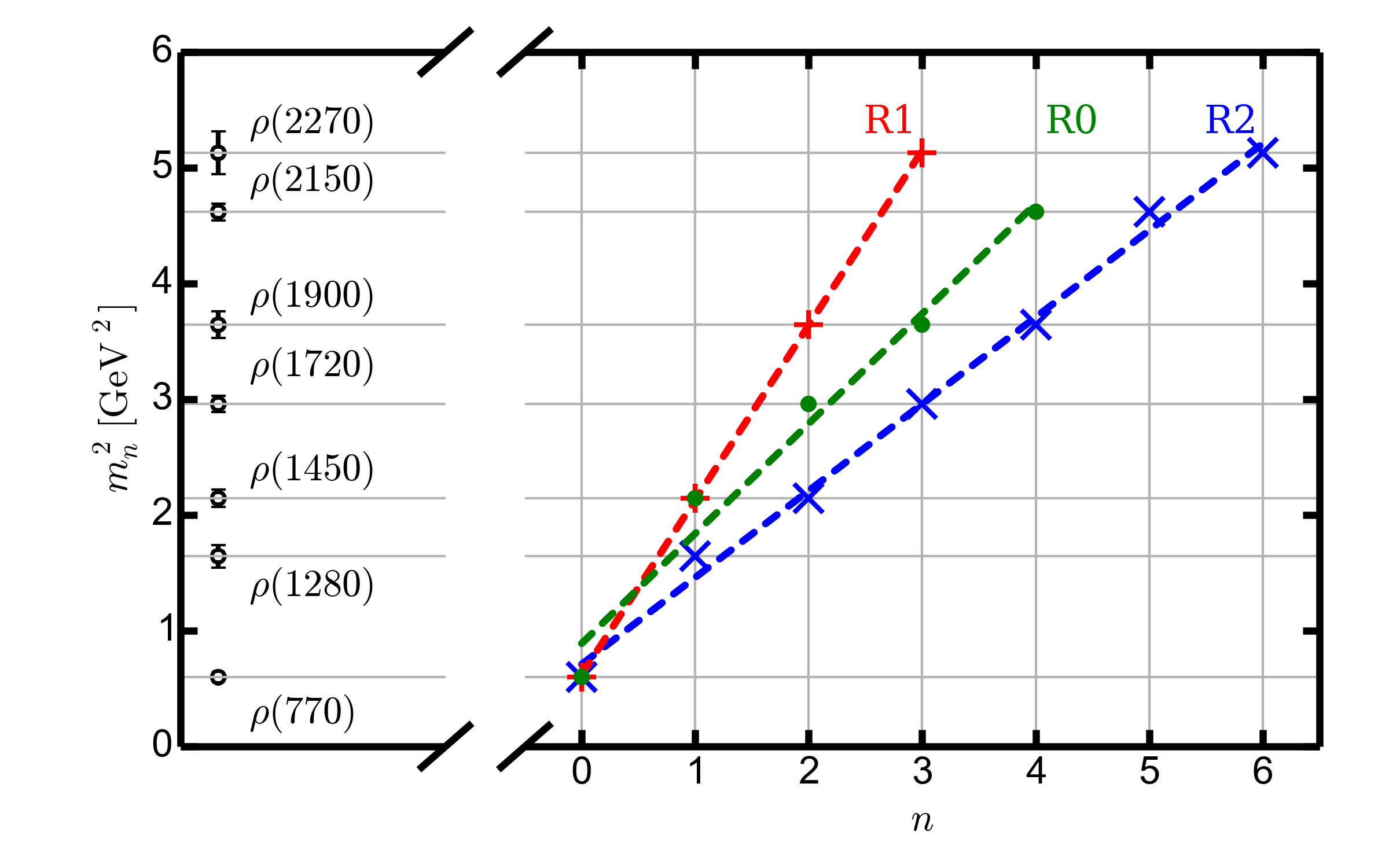} }
   \parbox{5.5cm}{\begin{tabular}{l | c c c}
  \hline
   & R0 & R1 & R2 \\
  \hline
  $\alpha$ [GeV$^2$] & 0.89 & 0.62 & 0.71 \\
  $\beta$ [GeV$^2$] & 0.95 & 1.51 & 0.75 \\
  \hline
  $-b$ & 0.253 & 2.540 & 0.056 \\
  $L^{-1}$ [MeV] & 487 & 614 & 433 \\
  \hline
  \end{tabular}}}
   \caption{Left: Various assignments of $J^{PC}=1^{--}$ $\rho$ meson masses squared \cite{pdg}, displayed on the l.h.s.,to radial quantum numbers $n$ (R0: green dots \cite{KKSS}, R1: red plusses \cite{Ebert}, R2: blue crosses \cite{BK}; for other options, cf.~\cite{masjuan}, and \cite{afonin} for the iso-scalar vector channel). The tentatively attributed Regge trajectories are depicted by dashed lines in the same colour code as the assignment symbols. Right: Parameters $\alpha$ and $\beta$ of $m_n^2 = \alpha +\beta n$ for the trajectories depicted in the left, as well as $b$ and $L^{-1}$ of the potential $U_0$ (\ref{u0bb}) determining $m_n^2$ from (\ref{schr}).} \label{abb-1}
  \end{figure}
  
To catch the physics and related scales of vector mesons (V) as a QCD incarnation we use the standard action \cite{KKSS} in Einstein frame
  \begin{eqnarray}
   S_V &\propto&  \int  \! \dd z\, \dd^4 x \,   \sqrt{g} F^2  \label{eq.2}
  \end{eqnarray}
and the ansatz for the infinitesimal line element squared in five-dimensional Riemann spacetime
  \begin{eqnarray}
   \dd s^2 &=&  e^{A(z)-\frac23 \Phi(z)} \left(f(z) \dd t^2 - \dd \vec x ^{\, 2} - \frac{\dd z^2}{f(z)} \right),  \label{ds}
  \end{eqnarray}
where $F^2$ is the squared field strength tensor of a $U(1)$ vector field, $A$ is a warp factor, $f$ the blackness function, and $\Phi$ denotes the dilaton field. The equation of motion follows, after some manipulations (cf.~\cite{Col09}), as one-dimensional Schr\"odinger type equation 
  \begin{equation} 
  \li(\Dp_{\xi}^2 -(U_T-m_n^2) \ri) \psi=0,  \label{schr}
  \end{equation}
where $\Dp_{\xi} = (1/f) \Dp_z$ and $U_T(z) = U_0 f^2 + \frac12 Sff'$ with
  \begin{equation}   
   U_0 (z) = \frac12 S' + \frac14 S^2, \quad
   S \equiv  \frac12 A' -\frac23 \Phi', \label{gs}
  \end{equation}
and a prime means derivative w.r.t.~$z$. Considering first the vacuum case, $f=1$ corresponding to zero temperature \cite{KKSS, Col09}, the ansatz (cf.~\cite{afonin} for instance)   
  \begin{equation}
   U_0= \frac34 \frac1{z^2} + 4 \frac{b}{L^2} + \li(\frac{z}{L}\ri)^2 \frac1{L^2} \label{u0bb}
  \end{equation}
delivers - with $L$ as scale parameter - the spectrum $m_n^2 = 4(n+b+1)/L^2$ of normalisable modes of (\ref{schr}) with $n=0,1,2,\ldots$ to be identified as radial quantum number. The table in Fig.~\ref{abb-1} lists optimal parameter choices which catch nicely several options for arranging the first few $\rho$ meson states as Regge trajectories (see left panel and upper part of the table in Fig.~\ref{abb-1}). \\
The requirement to have asymptotically an AdS spacetime, $\lim \limits_{z \to 0} A(z) = -2 \ln (z/L)$, gives constraints on $U_0$: The near-boundary expansions
  \begin{equation}
   S(z) = \frac1{z}  \sum \limits_{i=0}^{\un} s_i \frac{z^i}{L^i} \Ra  \frac12 S' + \frac14 S^2 = \frac1{z^2}  \sum \limits_{i=0}^{\un} u_i \frac{z^i}{L^i}
  \end{equation}
yield from (\ref{gs}, \ref{u0bb}) $u_2=0$ (which implies $b=0$) and $s_{2l}=0$ since $u_{2l+1}=0$ ($l=0,1,2,\ldots$). Inspection of the table in Fig.~\ref{abb-1} reveals that $b \neq0$ could be required for certain Regge trajectories. (The case $b=0$ recovers the original soft wall model \cite{KKSS}, where $\alpha=\beta=4/L^2$, corresponding to $1/L=487$~MeV, have been advocated.) A simple remedy is suppressing the impact of $b$ at the boundary by $U_0 \to \hat U_{0}(z) = \frac3{4z^2} + \frac{b}{L^2}(1-e^{-z^4/L^4}) + \frac{z^2}{L^4}$, since the potential $\hat U_{0}$ still generates approximately linear Regge trajectories. 

\section{Dilaton background}
Our next aim is to relate that useful $U_0$ to the dilaton profile $\Phi(z)$ and the warp factor $A(z)$, cf.~(\ref{gs}). To do so we consider the vector meson dynamics in the probe limit, i.e.~we seek for a suitable approach to $A$ and $\Phi$ without back reaction. One option is provided by the Einstein-dilaton model with action \cite{Kir1,Kir2,Gubser2}
  \begin{equation}
   S_{\Phi} = \frac{1}{k} \int \! \dd z \, \dd^4 x \sqrt{g}  \li[R -\frac{1}{2} \Dp_M \Phi \Dp^M \Phi - V(\Phi)\ri] , \label{1}
  \end{equation}
where the key quantity is the dilaton potential $V(\Phi)$, $R$ denotes the Einstein-Hilbert action and $k=1/(16\pi G_5)$, resulting with $a(z) \equiv \exp\{A(z)\}$ in the field equations
  \begin{eqnarray}
  f'' a +\frac32 a'f'&=& 0,  \label{allg0}\\
   a'' a -\frac32 {a'}^2+\frac13 {\Phi'}^2 a^2&=&0, \label{allg1} \\
   \li({a'}^2 -\frac16 {\Phi'}^2 a^2 \ri) f +\frac12 a'af' +\frac13 a^3 V &=&0. \label{allg2}
  \end{eqnarray}
Inserting the near-boundary expansions 
  \begin{equation}
   a(z) = \frac1{z^2} \sum \limits_{i=0}^{\un} a_i \frac{z^i}{L^i}, \quad
   \Phi(z) = \sum \limits_{i=0}^{\un} \phi_i \frac{z^i}{L^i},   \quad
   f(z) = \sum \limits_{i=0}^{\un} f_i \frac{z^i}{L^i}
  \end{equation}
one obtains $a_0=f_0=1$ and $a_{1,2,3}=f_{1,2,3}=0$ as well as $\phi_{0,1}=0$. In particular, $\phi_2 \neq 0$, i.e.~a quadratic dilaton profile in leading order is permitted iff $m_{\Phi}^2L^2=-4$. Such a profile is used in \cite{mei} as input to elucidate subsequently $V(\Phi)$ and the Regge trajectory. The case $m_{\Phi}^2L^2 < -4$ requires $\phi_2=0$, which is equivalent to $a_4=0$. A vanishing mass $m_{\Phi}$ is equivalent to a marginal operator and negative (positive) mass squared corresponds to a(n) (ir)relevant operator in the boundary theory, where one exploits the relations $\lim \limits_{\Phi\to0} V(\Phi) = -12/L^2 + m_{\Phi}^2 \Phi^2/2 + \ldots$ and $m_{\Phi}^2L^2=\Delta (\Delta-4)$ to characterise the dual operator to $\Phi$ by its conformal dimension $\Delta$. 
  \begin{figure}
  \cen{\includegraphics[scale=.53]{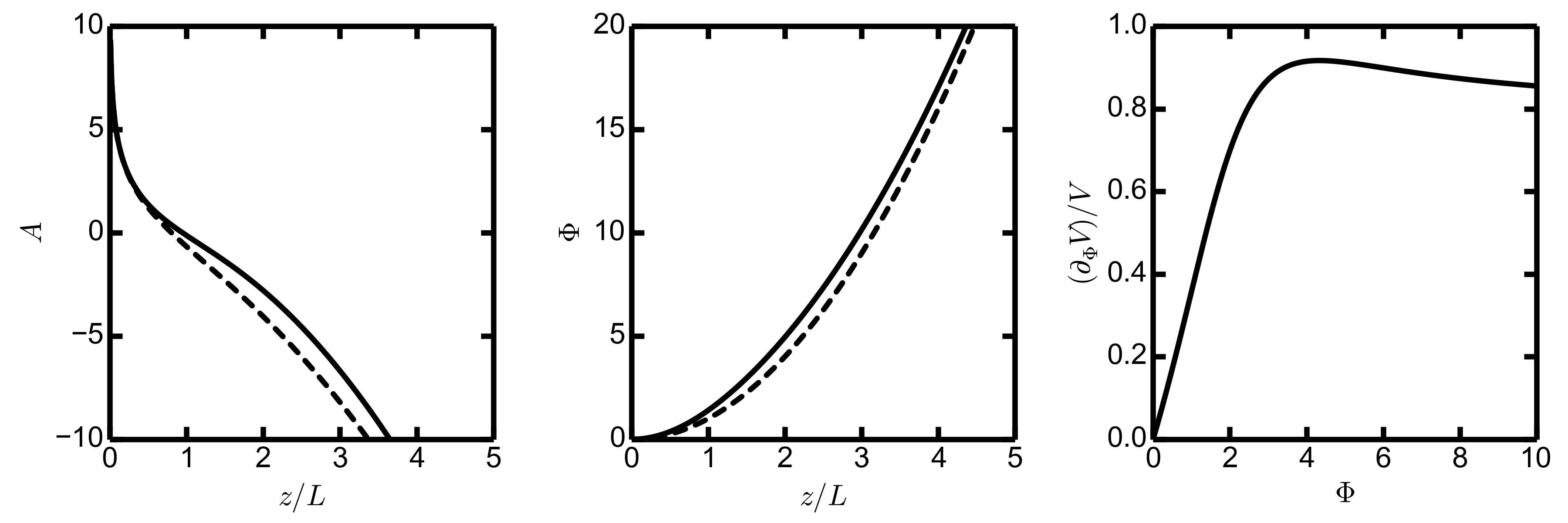}
  \caption{The warp factor $A$ (left) and the dilaton profile $\Phi$ (middle) as functions of $z/L$. The dilaton potential is represented by $(\Dp_{\Phi} V)/V$ as a function of $\Phi$ (right). The dashed curves are the profiles of the soft wall model.} \label{abb1}}
  \end{figure}
  
\hspace{-.64cm} Integrating the field equations (\ref{allg1}, \ref{allg2}) at $f=1$ with the side condition (\ref{gs}) means mapping $U_0(z)$ (or a particular hadron spectrum) on $V(\Phi)$. The corresponding profiles $A(z)$ and $\Phi(z)$ are exhibited in Fig.~\ref{abb1} (left and middle panels) as functions of $z/L$. For a comparison, $A^{\rm sw}=-2\ln(z/L)$ and $\Phi^{\rm sw} = z^2/L^2$, which are elements of the soft wall model \cite{KKSS}, are displayed too by dashed curves. In the displayed range, the soft wall profiles may serve as leading-order proxies of the solutions of the Einstein equations such to accommodate the Regge trajectory-determining potential $U_0$, albeit at $b=0$ and for the vacuum, $f=1$. This justifies a posteriori the ans\"atze in \cite{ich1, ich3} which explore certain deformations of the soft wall profiles with the goal to accommodate relevant QCD thermodynamics features. \\
The right panel in Fig.~\ref{abb1} exhibits $(\Dp_{\Phi} V)/V$ as a function of $\Phi$. With the aid of the adiabatic approximation \cite{Gubser2}, one can infer from $v_s^2= 1- \frac23 [(\Dp_{\Phi} V)/V]^2+ \ldots$ that for $(\Dp_{\Phi} V)/V > \sqrt{2/3}$ the velocity of sound, $v_s$, becomes imaginary, signalling the onset of a first-order phase transition, see also \cite{Yaresko}. In fact, solving the set of field equations (\ref{allg0}-\ref{allg2}) for $f(z,z_H) \leq 1$ and $f(z=z_H,z_H)=0$, i.e.~admitting a horizon at $z_H$, with the same potential $V(\Phi)$ which is found for the above prescribed potential $U_0$, one sees that the temperature, $T(z_H)=-\Dp_zf(z,z_H) \mid _{z=z_H} /(4\pi)$ has a global minimum at $z_H^{\min} = 1.2/L$ with $T(z_H^{\min})=116$~MeV for $1/L=687$~MeV. As extensively discussed in \cite{Kir1, Kir2} this refers to a first-order phase transition with transition temperature $T_{\rm c} = 1.02\, T(z_H^{\min})$. 

  \begin{figure}
   \cen{\includegraphics[scale=.53]{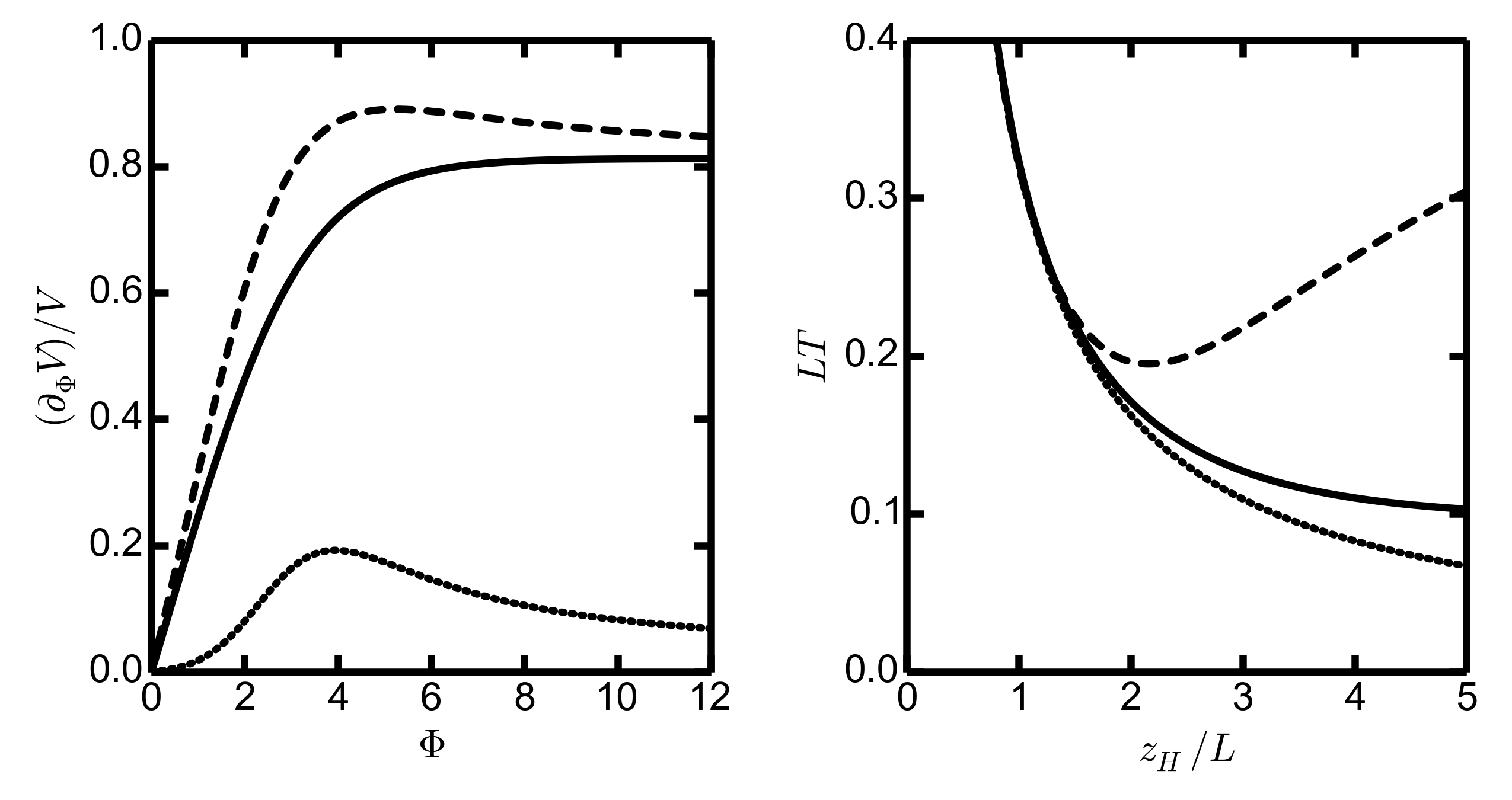}
   \caption{$(\Dp_{\Phi}V)/V$ as a function of $\Phi$ (left panel) for $L^2V(\Phi,p)  = -12\,  _0F_1^2\left(\frac{1}{2p},\frac{\Phi^2}{24}\right) +\frac{p^2}{2} \Phi^2 \,  _0F_1^2\left(\frac{2p+1}{2p},\frac{\Phi^2}{24}\right)$ and the resulting temperature $T$ as a function of the horizon position $z_H$ (right panel). Dashed curves: $p=1.5$, $T(z_H)$ has a global minimum (which induces a first-order phase transition; approximate Regge trajectories  within $U_0$ at $T=0$ but not within $U_T$); dotted curves: $p=0.5$, $T(z_H)$ has an inflection point (which points to a cross over; at $T=0$, the potentials $U_0$ and $U_T$ do not allow for meson states at all); solid curves: $p=1$, delineation case, where $T(z_H)$ asymptotes to a constant value at large values of $z_H$. The temperature independent potentials $V(\Phi,p)$ and $U_0$ follow here from (\ref{allg1}, \ref{allg2}) at $f=1$ by the ansatz $\Phi =(z/L)^p$.} \label{abb3}}
  \end{figure}

\hspace{-.6cm} It seems conceivable that a pure gravity-one--component dilaton model has a too restricted field content to mimic QCD thermodynamics and vacuum hadron (at least, vector meson) spectra by one unique potential $V(\Phi)$. For instance, considering a special ansatz for $V(\Phi)$ we find that either proper Regge trajectories and a first-order phase transition \underline{or} QCD thermodynamics with a cross over and none discrete hadron states are realised in such a scenario, see Fig.~\ref{abb3}. (This supplements the assessment in \cite{nitti} for the glue sector based on (\ref{1}): a gapped, discrete spectrum at $T=0$ facilitates a first-order transition at $T>0$.) A further issue is that holographic models face often a too low temperature of hadron melting \cite{9,10,11,12}. The above value of $T_{\rm c}$ is a reminiscence of that insanity, as it is significantly below the QCD cross-over temperature of $\mathcal{O}(150~{\rm MeV})$.

\section{Conclusions}
The constraints from the mass spectra of radial $\rho$ meson excitations forming Regge trajectories can be used to construct the dilaton potential $V(\Phi)$ within an Einstein-dilaton background model by solving the field equations at zero temperature (vacuum). However, the same $V(\Phi)$ generates a first-order (Hawking-Page type) phase transition in the case of non-zero temperatures. That kind of phase structure stands in conflict with 2+1 flavour QCD with physical quark masses \cite{Borsanyi2014, Bazavov2014}, however, it is in line with expectations of 2+1 flavour QCD near the chiral limit, where firm quantitative results are still lacking \cite{Bazavov2017}. Hence, the strategy of our following work will be seeking other avenues to the gravity background and dilaton potential such to accommodate the vacuum hadron spectrum and QCD thermodynamics at the same time, as envisaged in the ansatz trials \cite{ich1,ich3,ich2}, as a prerequisite to uncover the $T$-$\mu$ plane (cf.~\cite{Knaute, critelli} for thermodynamics aspects and \cite{itrakis, wolf} for the general layout) in the FAIR relevant region by holographic means.

\ack{Useful discussions with Profs.~Mei Huang and Jorge Noronha are gratefully acknowledged. The work is supported by Studienstiftung des deutschen Volkes. }

\vspace{0.5cm}
 
\hrule


\vspace{0.5cm}

\begin{thebibliography}{99}
 \bibitem{tetyana} T. Galatyuk et al., Eur. Phys. J. A 52, 131 (2016).
 \bibitem{Erdmenger} M. Ammon, J. Erdmenger, Gauge/Gravity duality, Cambridge University Press (2015).
 \bibitem{Kir1} U. G\"ursoy, E. Kiritsis, JHEP 0802, 032 (2008).
 \bibitem{Kir2} U. G\"ursoy, E. Kiritsis, F. Nitti, JHEP 0802, 019 (2008).
 \bibitem{KKSS} A. Karch, E. Katz, D. T. Son, M. A. Stephanov, Phys. Rev. D 74, 015005 (2006).
 \bibitem{Col09} P. Colangelo, F. Giannuzzi, S. Nicotri, Phys. Rev. D 80, 094019 (2009). %Holographic Approach to Finite Temperature QCD: %The Case of Scalar Glueballs and Scalar Mesons 
 \bibitem{pdg}  C. Patrignani et al. (Particle Data Group), Chin. Phys. C 40, 100001 (2016).
 \bibitem{Ebert} D. Ebert, R. N. Faustov, V. O. Galkin, Phys. Rev. D 79, 114029 (2009).
 \bibitem{BK} S. P. Bartz, J. I. Kapusta, Phys. Rev. D 90, 074034 (2014).
 \bibitem{masjuan} P. S. Masjuan, E. Arriola, W. Broniowski, Phys. Rev. D 85, 094006 (2012).
 \bibitem{afonin} S. S. Afonin, Phys. Lett. B 719, 399 (2013).
 \bibitem{Gubser2} S. S. Gubser, A. Nellore, Phys. Rev. D 78, 086007 (2008).
 \bibitem{mei} D. Li, M. Huang, JHEP 1311, 088  (2013).
 \bibitem{ich1} R. Z\"ollner, B. K\"ampfer, Phys. Rev. C 94, 045205 (2016).
 \bibitem{ich3} R. Z\"ollner, B. K\"ampfer, Eur. Phys. J. A 53, 139 (2017).
 \bibitem{nitti} U. G\"ursoy, E. Kiritsis, L. Mazzanti, F. Nitti, JHEP 05, 033 (2009).
 \bibitem{9} {M. Fujita, T. Kikuchi, K. Fukushima, T. Misumi, M. Murata, {Phys. Rev.} D {81}, 065024 (2010).}
 \bibitem{10} {T. Ishii, S. Kinoshita, K. Murata, N. Tanahashi, {JHEP} {1404}, 099 (2014).}
 \bibitem{11} {N. R. F. Braga, M. A. Martin Contreras, S. Diles, {Eur. Phys. J.} C {76}, 598 (2016).}
 \bibitem{12} {S. P. Bartz, T. Jacobson, {Phys. Rev.} D {94}, 075022 (2016).}
 \bibitem{Borsanyi2014} S. Bors$\rm \acute{a}$nyi et al., Phys. Lett. B 370, 99 (2014).
 \bibitem{Bazavov2014} A. Bazavov et al., Phys. Lett. B 737, 210 (2014).
 \bibitem{Bazavov2017} A. Bazavov et al., Phys. Rev. D 95, 074505 (2017).
 \bibitem{Yaresko} R. Yaresko, B. K\"ampfer, Phys. Lett. B 747, 36 (2015).
 \bibitem{ich2} R. Z\"ollner, F. Wunderlich, B. K\"ampfer, arXiv:1611.04124 [hep-th] (2016).
 \bibitem{Knaute} J. Knaute, R. Yaresko, B. K\"ampfer, arXiv:1702.06731 [hep-th] (2017).
 \bibitem{critelli} R. Critelli et al., arXiv:1706.00455 [hep-th] (2017).
 \bibitem{itrakis} I. Iatrakis, E. Kiritsis, A. Paredes, JHEP 1011, 123 (2010).
 \bibitem{wolf} O. De Wolfe, S. S. Gubser, C. Rosen, Phys. Rev. D 83, 086005 (2011).
\end{thebibliography}
\end{document}